\documentclass[showpacs,preprintnumbers,amsmath,amssymb]{revtex4}
\usepackage{graphicx}

\begin{document}

\title{Statefinder Parameters for Five-Dimensional Cosmology}

\author{Baorong Chang\footnote{changbaorong@student.dlut.edu.cn},
Hongya Liu\footnote{Corresponding author: hyliu@dlut.edu.cn},
Lixin Xu and Chengwu Zhang}

\address{School of Physics \& Optoelectronic Technology, Dalian
University of Technology, Dalian, 116024, P. R. China}


\begin{abstract}
We study the statefinder parameter in the five-dimensional big
bounce model, and apply it to differentiate the attractor
solutions of quintessence and phantom field. It is found that the
evolving trajectories of these two attractor solutions in the
statefinder parameters plane are quite different, and that are
different from the statefinder trajectories of other dark energy
models.
\end{abstract}

\maketitle



\section{Introduction}

Recent observations of Cosmic Microwave Background (CMB)
anisotropies indicate that the universe is flat and the total energy
density is very close to the critical one with $\Omega
_{total}\simeq 1$ \cite{CMB}. Meanwhile, observations of high
redshift type Ia supernovae\cite{Ia} reveal the speeding up
expansion of our universe and the surveys of clusters of galaxies
show that the density of matter is very much less than the critical
density\cite{SDSS}. These three tests nicely complement each other
and indicate that the dominated component of the present universe is
dark energy. The Wilkinson Microwave Anisotropy Probe (WMAP)
satellite experiment tells us that dark energy, dark matter, and the
usual baryonic matter occupy about $73\%$, $23\%$, and $4\%$ of the
total energy of the universe, respectively. The accelerated
expansion of the present universe is attributed to the dark energy
whose essence is quite unusual and there is no justification for
assuming that it resembles known forms of matter or energy.
Candidates for dark energy have been widely studied and focus on the
cosmological constant $\Lambda$ \cite{sahni9904398,cosmological
constant,cosmological constant1,cosmological constant2,cosmological
constant3} with $w=-1$, a dynamically evolving scalar field
(quintessence)
\cite{quintessence1,quintessence11,quintessence12,quintessence13,quintessence2,quintessence21,quintessence22}
with $w>-1$ and phantom \cite{phantom,phantom1,phantom2,phantom3}
with $w<-1$. The fine tuning problem is considered as one of the
most important issues for dark energy models and a good model should
limit the fine tuning as much as possible. The dynamical attractors
of the cosmological system have been employed to make the late time
behaviors of the model insensitive to the initial conditions of the
field and thus alleviate the fine tuning problem, which has been
studied in many quintessence models
\cite{quattractor1,quattractor11,quattractor2,quattractor0}. In
phantom field, this problem has been studied in
\cite{phattractor1,phattractor11,phattractor12,phattractor2}. There
are many dark energy models are constructed for interpreting the
cosmic acceleration and solving the fine tuning problem. In order to
differentiate these models, a diagnostic proposal that makes use of
parameter pair $\{r,s\}$, the so-called statefinder, was introduced
by Sahni et al.\cite{statefinder} and defined as follows:
\begin{equation}
r\equiv\frac{\dddot{a}}{aH^{3}},\qquad
s\equiv\frac{r-1}{3(q-1/2)}.\label{sf}
\end{equation}
Here $q$ is the deceleration parameter. The statefinder is a
"geometrical" diagnostic in the sense that it depends on the
expansion factor and hence on the metric describing space-time.
Since different cosmological models involving dark energy exhibit
qualitatively different evolution trajectories in the $s-r$ plane,
this statefinder diagnostic can differentiate various kinds of dark
energy models. For the spatially flat LCDM cosmological model, the
statefinder parameters correspond to a fixed point $\{r=1,s=0\}$. By
far some models, including the cosmological constant, quintessence,
phantom, quintom, the Chaplygin gas, braneworld models, holographic
models, interacting and coupling dark energy models
\cite{grqc0311067,statefinder,models for statefinder,models for
statefinder1,models for statefinder2,models for statefinder3,models
for statefinder4,models for statefinder5,models for statefinder6},
have been successfully differentiated in the standard 4D FRW model.

In Kaluza-Klein theories as well as in brane world scenarios, our 4D
universe is believed to be embedded in a higher-dimensional
manifold. In this paper, we consider a class of five-dimensional
cosmological model which as an alternative candidate to the standard
4D FRW model, has been discussed by many authors
\cite{APJ,ourwork,ourwork1,ourwork2,ourwork3,ourwork4,ourwork5,ourwork6,ourwork7,ourwork8,ourwork9,ourwork10}.
Instead of the Big Bang singularity of the standard model, this 5D
cosmological model is characterized by a "Big Bounce", which
corresponds to a finite and minimal size of the universe. Before the
bounce the universe contracts, and after the bounce it expands. This
model is 5D Ricci-flat, implying that it is empty viewed from 5D.
However, as is known from the induced matter theory
\cite{wesson,STM}, 4D Einstein equations with matter could be
recovered from 5D equations in apparent vacuum. This approach is
guaranteed by Campbell's theorem that any solution of the Einstein
equations in N-dimensions can be locally embedded in a Ricci-flat
manifold of (N+1)-dimensions \cite{campbell}. The purpose of this
paper is to study the statefinder parameter in the five-dimensional
bounce model, and apply it to contrast the attractor solutions of
quintessence and phantom field respectively. In section II, we
introduce the scalar field in the 5D bounce cosmological solutions
and the 5D quintessence and phantom model for dark energy, and
deduce the statefinder parameter in 5D universe. Section III and IV
are to apply the statefinder parameters to analyze the attractor
solutions of quintessence and phantom respectively in 5D universe.
Section V is a short discussion.

\section{Statefinder Parameter in the 5D Model}

An exact 5D cosmological solution was given firstly by Liu and
Mashhoon in \cite{Mashhoon} and restudied by Liu and Wesson in
\cite{APJ}. This solution reads
\begin{eqnarray}
dS^{2}&=& B^{2}(t,y)dt^{2}-A^{2}(t,y)\left[\frac{dr^{2}}{1-kr^{2}}
+r^{2}(d\theta ^{2}+\sin ^{2}\theta d\phi ^{2})\right] -dy^{2},  \nonumber \\
B&=&\frac{1}{\mu}\frac{\partial A}{\partial t}\equiv \frac{\dot{A}}{\mu}
\label{5Dmatric} \\
A^{2}&=&(\mu ^{2}+k)y^{2}+2\nu y+\frac{\nu ^{2}+K}{\mu ^{2}+k},  \nonumber
\end{eqnarray}
where $\dot{A}=(\partial /\partial t)A$, $\mu =\mu (t)$ and $\nu
=\nu (t)$ are two arbitrary functions of $t$ , $k$ is the 3-D
curvature index $(k=\pm 1,0)$ and $K$ is a constant. This solution
satisfies the 5D vacuum equation $R_{AB}=0$, with the three
invariants being
\begin{equation}
I_{1}\equiv R=0,\qquad\qquad I_{2}\equiv
R^{AB}R_{AB}=0,\qquad\qquad I_{3}\equiv R^{ABCD}R_{ABCD}=
\frac{72K^{2}}{A^{8}}.  \label{b}
\end{equation}
So $K$ determines the curvature of the 5D manifold.

The 5D line element in (\ref{5Dmatric}) contains the 4D one,
\begin{equation}
ds^{2}=g_{\mu \nu }dx^{\mu }dx^{\nu }=B^{2}dt^{2}-A^{2}\left[
\frac{dr^{2}}{1-kr^{2}}+r^{2}(d\theta ^{2}+\sin ^{2}\theta d\phi
^{2})\right] .  \label{4Dline}
\end{equation}
Using this 4D metric we can calculate the 4D Einstein tensor by $
^{(4)}G_{\nu }^{\mu }\equiv ^{(4)}\smallskip R_{\nu }^{\mu
}-\delta _{\nu }^{\mu }$ $^{(4)}R/2$. Its non-vanishing components
are
\begin{eqnarray}
^{(4)}G_{0}^{0} &=&\frac{3\left( \mu ^{2}+k\right) }{A^{2}},  \nonumber \\
^{(4)}G_{1}^{1} &=&^{(4)}G_{2}^{2}=^{(4)}G_{3}^{3}=\frac{2\mu \dot{\mu}}{A%
\dot{A}}+\frac{\mu ^{2}+k}{A^{2}}. \label{4Ccomponents}
\end{eqnarray}
Generally speaking, the Einstein tensor in (\ref{4Ccomponents})
can give a 4D effective or induced energy-momentum tensor
$^{(4)}T_{\nu }^{\mu }$ via $ ^{(4)}G_{\nu }^{\mu }=\kappa
^{2\,(4)}T_{\nu}^{\mu }$ with $ \kappa ^{2}=8\pi G$.

In previous works
\cite{APJ,ourwork,ourwork1,ourwork2,ourwork3,ourwork4,ourwork5,ourwork6,ourwork7,ourwork8,ourwork9,ourwork10},
this energy-momentum tensor was assumed to be a perfect fluid with
density $\rho $ and pressure $p$, plus a variable cosmological term
$\Lambda $, and the results shows that this assumption works well.
However, there are more and more helpful candidates for dark energy
seem to be the scalar field models such as the quintessence and
phantom. So, in Ref. \cite{phattractor2}, the authors constructed
the model that the 4D induced energy-momentum tensor consists of two
parts:
\begin{eqnarray}
^{(4)}T_{\mu \nu }&=&T_{\mu \nu }^{m}+T_{\mu \nu }^{\phi },  \nonumber \\
T_{\mu \nu }^{m}&=&(\rho _{m}+p_{m})u_{\mu }u_{\nu }-p_{m}g_{\mu
\nu},
\label{fluid} \\
T_{\mu \nu }^{\phi }&=&{\varepsilon}{\partial}_{\mu }\phi
\partial_{\nu }\phi-g_{\mu \nu }[\frac{1}{2}{\varepsilon}g^{\mu
\nu }\partial_{\mu }\phi\partial_{\nu }\phi-V(\phi )], \nonumber
\end{eqnarray}
where $T_{\mu\nu }^{m}$ represents a perfect fluid and $T_{\mu\nu
}^{\phi }$ represents a scalar field with $\varepsilon=\pm1$. For
$\varepsilon=+1$, $T^{\phi}_{\mu\nu}$ is regular and it represents
the quintessence model. For $\varepsilon=-1$, the scalar field has
a negative dynamic energy and $T^{\phi}_{\mu\nu}$ represents the
phantom model. And, we assume that each of the two components
$T^{m}_{\mu\nu}$ and $T^{\phi}_{\mu\nu}$ conserves independently
just as in most of the 4D dark energy models. \\
Here, we should notice that the coordinate time $t$ is not the
proper time in the solutions (\ref{5Dmatric}), and, generally, one
cannot transform it to the proper time without changing the form
of the 5D metric. However, on a given hypersurface $y=const$, the
proper time $\tau$ relates the coordinate time $t$ via
$d\tau=Bdt$. So, in this 5D model, the Hubble parameter $H$, the
deceleration parameter $q$, and the statefinder parameter
(\ref{sf}) introduced in Ref. (\cite{statefinder}) should be given
as follows:
\begin{eqnarray}
H&=&\frac{\dot{A}}{AB}, \nonumber \\
q&=&-\frac{A\ddot{A}}{\dot{A}^{2}}+\frac{A\dot{B}}{B\dot{A}}, \nonumber\\
r&=&\frac{A^{2}\dddot{A}}{\dot{A}^{3}}-3\frac{A^{2}\ddot{A}\dot{B}}{\dot{A}^{3}B}
-\frac{A^{2}\ddot{B}}{\dot{A}^{2}B}+3\frac{A^{2}\dot{B}^{2}}{\dot{A}^{2}B^{2}}, \nonumber \\
s&=&\frac{r-1}{3(q-1/2)}. \label{Hrsq}
\end{eqnarray}
Using the relations (\ref{4Ccomponents})-(\ref{fluid}) and the
conservation laws $T^{(m)\nu}_{\mu;\nu}=0$ and
$T^{(\phi)\nu}_{\mu;\nu}=0$, we can obtain the 4D Einstein
equation and the equations of motion for the scalar field:

\begin{eqnarray}
\dot{\rho}_{m}&+&3\frac{\dot{A}}{A}(\rho_{m}+p_{m})=0,  \nonumber \\
\ddot{\phi}+(3\frac{\dot{A}}{A}&-&\frac{\dot{B}}{B})\dot{\phi}+{\varepsilon}B^{2}\frac{dV}{d\phi}=0,  \nonumber \\
H^{2}&+&\frac{k}{A^{2}}=\frac{\kappa^{2}}{3}(\rho_{m}+\rho_{\phi}),  \nonumber \\
\dot{H}=-&\frac{\kappa^{2}}{2}&B(\rho_{m}+p_{m}+\rho_{\phi}+p_{\phi}).\label{motion}
\end{eqnarray}
where
\begin{equation}
\rho_{\phi}\equiv\frac{1}{2}\varepsilon\frac{\dot{\phi}^{2}}{B^{2}}+V(\phi),\qquad\qquad
p_{\phi}\equiv\frac{1}{2}\varepsilon\frac{\dot{\phi}^{2}}{B^{2}}-V(\phi).\label{density}
\end{equation}
are the energy density and pressure of the scalar field
respectively. The potential $V(\phi)$ is assumed to be
exponentially dependent on $\phi$ by
$V(\phi)=V_{0}\exp(-\lambda\kappa\phi)$. $\rho_{m}$ and $P_{m}$
are the energy density and pressure of the perfect fluid
respectively, and $P_{m}=(\gamma_{m}-1)\rho_{m}$.\\
The equation of state parameter for the scalar field is found to
be:
\begin{equation}
w_{\phi}=\frac{p_{\phi}}{\rho_{\phi}}=\frac{\varepsilon\dot{\phi}^{2}-2B^{2}V(\phi)}{\varepsilon\dot{\phi}^{2}+2B^{2}V(\phi)}.\label{w}
\end{equation}

\section{Statefinder Parameter for 5D Attractor Solution of Quintessence Model}

\begin{figure}
\begin{center}
\includegraphics[width=2.4in]{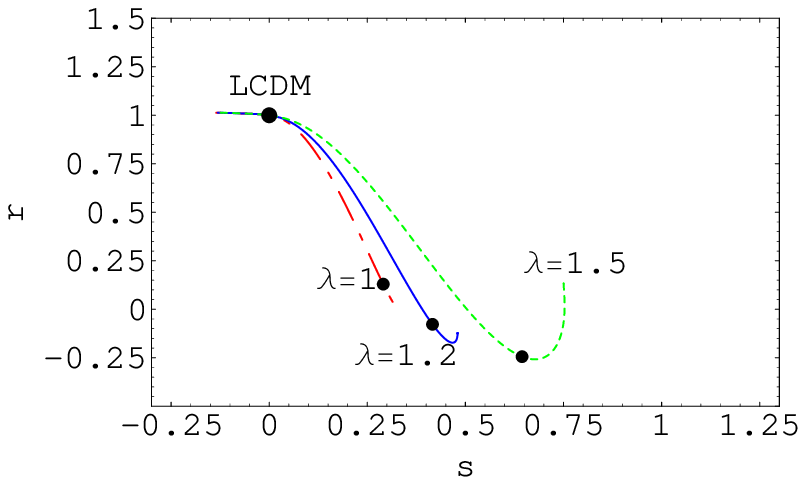}\includegraphics[width=2.4in]{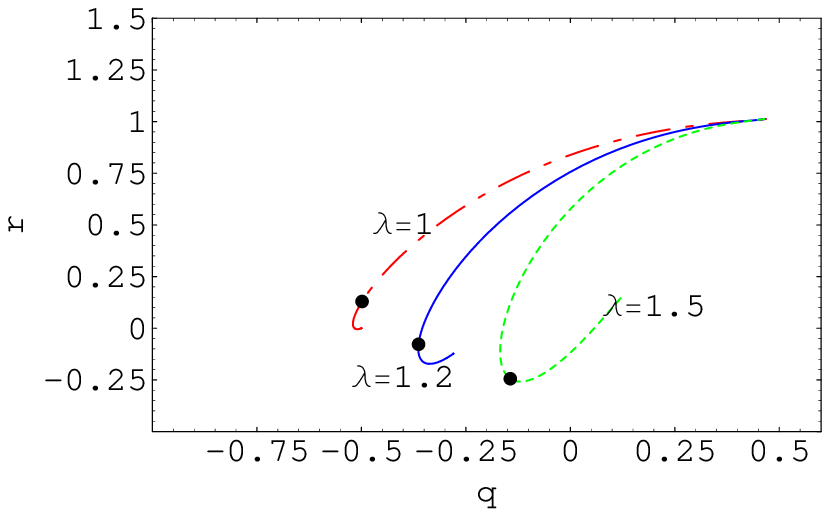}
\end{center}
\caption{The left figure is $s-r$ diagram of attractor solution in
quintessence model. The right figure is $q-r$ diagram of attractor
solution in quintessence model. The curves evolve in the variable
interval $N\in[-2,10]$. Selected curves for $\gamma_{m}=1$,
$\lambda=1$, $1.2$ and $1.5$ respectively. Dots locate the current
values of the statefinder parameters.}\label{sr1}
\end{figure}

\begin{figure}
\begin{center}
\includegraphics[width=2.4in]{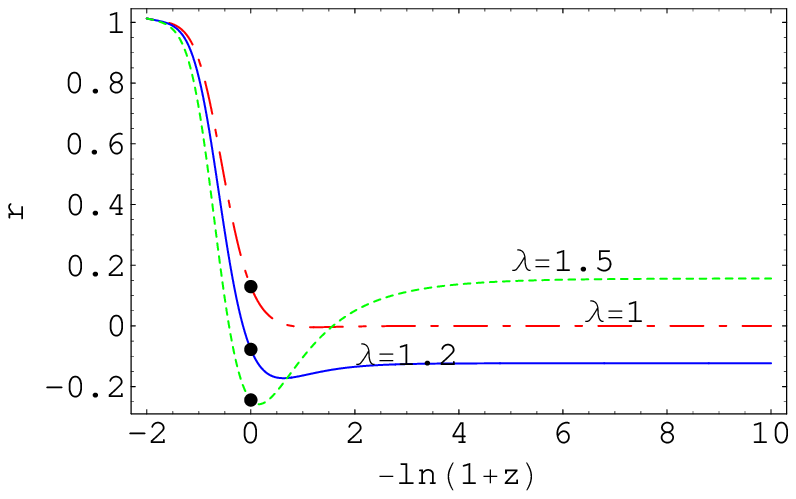}\includegraphics[width=2.4in]{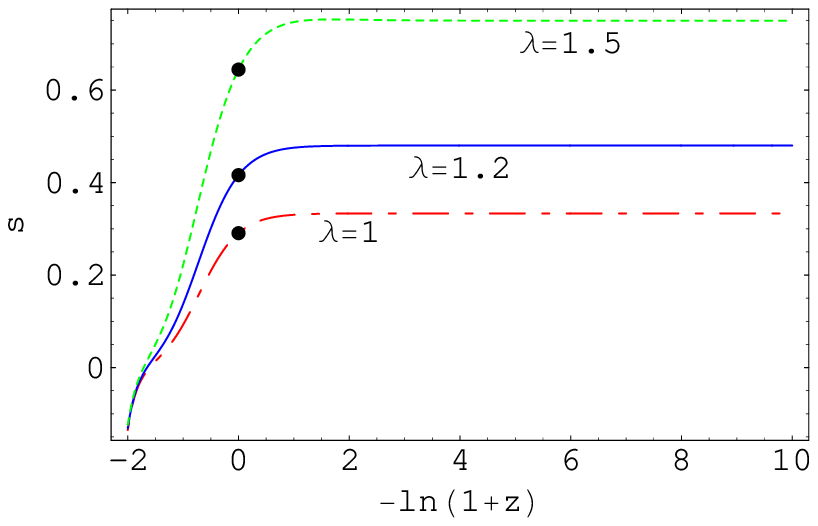}\includegraphics[width=2.4in]{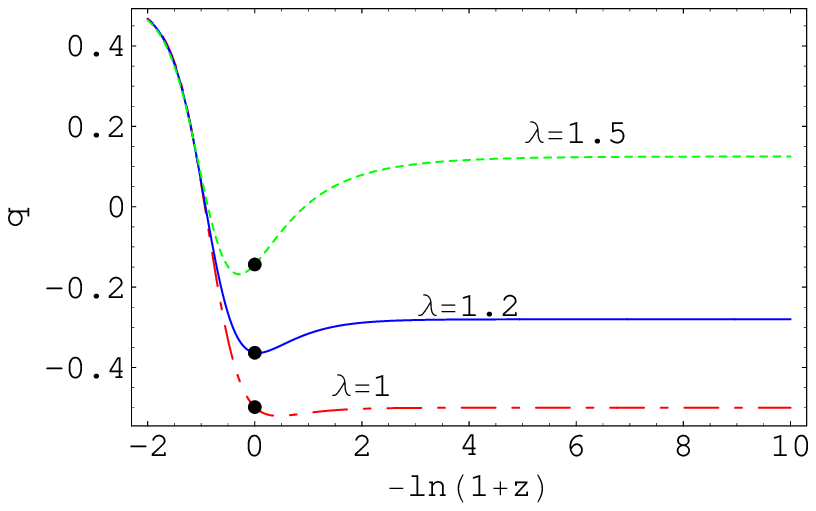}
\end{center}
\caption{The figures are the statefinder parameters and
deceleration parameter versus redshift diagram of the attractor
solution in quintessence model. The left figure is $-\ln(1+z)$ ---
$r$ diagram, the middle figure is $-\ln(1+z)$ --- $s$ diagram and
the right figure is $-\ln(1+z)$
--- $q$ diagram. The curves evolve in the variable interval $N\in[-2,10]$. Selected curves for $\gamma_{m}=1$,
$\lambda=1$, $1.2$ and $1.5$ respectively. Dots locate the current
values of the statefinder parameters.}\label{nr1}
\end{figure}

The quintessence model is the scenario of equations
(\ref{motion})-(\ref{w}) with $\varepsilon=+1$, which has been
discussed in previous work \cite{quattractor0}. As in Ref.
\cite{quattractor2}, we define the $x$ and $y$ in a
plane-autonomous system as:

\begin{equation}
x=\frac{\kappa\dot{\phi}}{\sqrt{6}BH}, \qquad\qquad
y=\frac{\kappa\sqrt{V}}{\sqrt{3}H}.  \label{xy}
\end{equation}
For a spatially flat universe $(k=0)$, we find that the evolution
equation for $x$ and $y$ are of the same form as in Ref.
\cite{quattractor2},
\begin{eqnarray}
x^{'}&=&-3x+\lambda\sqrt{\frac{3}{2}}y^{2}+\frac{3}{2}x[2x^{2}+\gamma_{m}(1-x^{2}-y^{2})],  \nonumber \\
y^{'}&=&-\lambda\sqrt{\frac{3}{2}}xy+\frac{3}{2}y[2x^{2}+\gamma_{m}(1-x^{2}-y^{2})],\label{quxy}
\end{eqnarray}
where a prime denotes a derivative with respect to the logarithm
of the scale factor, $N=\ln A$. And we can get the densities of
two components:
\begin{equation}
\Omega_{m}=\frac{\kappa^{2}\rho_{m}}{3H^{2}}=1-x^{2}-y^{2},\qquad\qquad
\Omega_{\phi}=\frac{\kappa^{2}\rho_{\phi}}{3H^{2}}=x^{2}+y^{2}.\label{qudensity}
\end{equation}
and the equation of state parameter for quintessence is:
\begin{equation}
w_{\phi}=\frac{x^{2}-y^{2}}{x^{2}+y^{2}}.\label{quw}
\end{equation}
The statefinder parameters and deceleration parameter for this
quintessence model are:
\begin{eqnarray}
r&=&(\frac{9}{2}\gamma_{m}^{2}-\frac{9}{2}\gamma_{m}+1)(1-x^{2}-y^{2})-3\sqrt{6}{\lambda}xy^{2}+10x^{2}+y^{2},  \nonumber \\
s&=&\frac{(\frac{9}{2}\gamma_{m}^{2}-\frac{9}{2}\gamma_{m}+1)(1-x^{2}-y^{2})-3\sqrt{6}{\lambda}xy^{2}+10x^{2}+y^{2}-1}{3[(\frac{3}{2}\gamma_{m}-1)(1-x^{2}-y^{2})+2x^{2}-y^{2}-\frac{1}{2}]},  \nonumber \\
q&=&(\frac{3}{2}\gamma_{m}-1)(1-x^{2}-y^{2})+2x^{2}-y^{2},\label{rsxy}
\end{eqnarray}
There are two late-time attractor solutions for (\ref{quxy}),
which have been found in Ref. \cite{quattractor0}. One is the
scaling solution which has been discussed by the authors in Ref.
\cite{quattractor0}, another is the late-time attractor solution
dominated by the scalar field:
\begin{equation}
x=\lambda /\sqrt{6},\qquad\qquad
y=\sqrt{1-\lambda^{2}/6},\qquad\qquad with\qquad\lambda^{2}<6
\label{qusolution}
\end{equation}
In the following we will discuss the statefinder parameters for
the attractor solution of quintessence model. In Fig. \ref{sr1} we
plot the evolution of the statefinder pairs $\{r,s\}$ and
$\{r,q\}$ for the attractor solution of quintessence. The plot is
for variable interval $N\in[-2,10]$, and the selected evolution
trajectories of $r(s)$ and $r(q)$ correspond to $\gamma_{m}=1$,
$\lambda=1$, $1.2$ and $1.5$ respectively. From the evolution
trajectory of $r(s)$ we can see that the curves pass through the
fixed point LCDM in the past, while the distance from the curves
to LCDM scenario is somewhat far in the future. The statefinder
pair $\{r,s\}$ lies in the regions $s>0$, $r<1$, which is differ
from other quintessence model. The diagram of $-ln(1+z)$
--- $r$ and $-ln(1+z)$ --- $s$ in Fig. \ref{nr1} show that the big
deviation between statefinder parameter and the LCDM scenario has
been caused with the increase of the parameter $\lambda$ in the
potential $V(\phi)$. From the diagram $-ln(1+z)$ --- $q$ we can
see that the deceleration parameter $q$ satisfies $q<0$ at
present. The Fig. \ref{quw} is the curves of the equation of state
versus redshift with different $\lambda$, which indicates that the
universe is accelerating without appearance of the big rip in the
future.

\begin{figure}
\begin{center}
\includegraphics[width=2.4in]{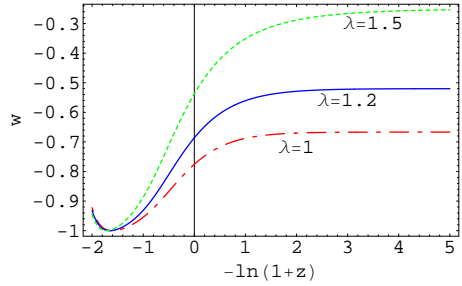}
\end{center}
\caption{The figures are the equation of state of the quintessence
model $w$ versus redshift.  The curves evolve in the variable
interval $N\in[-2,5]$. Selected curves for $\gamma_{m}=1$,
$\lambda=1$, $1.2$ and $1.5$ respectively.}\label{quw}
\end{figure}

\section{Statefinder Parameter for 5D Attractor Solution of Phantom Model}

The phantom model is the scenario of equations
(\ref{motion})-(\ref{w}) with $\varepsilon=-1$, which has been
discussed in previous work \cite{phattractor2}.  As in Ref.
\cite{phattractor2}, we define the dimensionless variables:
\begin{equation}
x=\frac{\kappa\dot{\phi}}{\sqrt{6}BH}, \qquad\qquad
y=\frac{\kappa\sqrt{V}}{\sqrt{3}H}.  \label{phxy}
\end{equation}
For a spatially flat universe $(k=0)$, we find that the evolution
equation for $x$ and $y$ in the phantom model are:
\begin{eqnarray}
x^{'}&=&\frac{3}{2}x[-2x^{2}+\gamma_{m}(1+x^{2}-y^{2})]-3x-\lambda\sqrt{\frac{3}{2}}y^{2},  \nonumber \\
y^{'}&=&\frac{3}{2}y[-2x^{2}+\gamma_{m}(1+x^{2}-y^{2})]-\lambda\sqrt{\frac{3}{2}}xy.\label{phphaseplane}
\end{eqnarray}
where a prime denotes a derivative with respect to the logarithm
of the scale factor, $N=\ln A$. And we can get the densities of
two components:
\begin{equation}
\Omega_{m}=\frac{\kappa^{2}\rho_{m}}{3H^{2}}=1+x^{2}-y^{2},\qquad\qquad
\Omega_{\phi}=\frac{\kappa^{2}\rho_{\phi}}{3H^{2}}=-x^{2}+y^{2}.\label{phdensity}
\end{equation}
and the equation of state parameter for phantom is:
\begin{equation}
w_{\phi}=\frac{x^{2}+y^{2}}{x^{2}-y^{2}}.\label{phw}
\end{equation}
The statefinder parameters and deceleration parameter for this
phantom model are:
\begin{eqnarray}
r&=&(\frac{9}{2}\gamma_{m}^{2}-\frac{9}{2}\gamma_{m}+1)(1+x^{2}-y^{2})-3\sqrt{6}{\lambda}xy^{2}-10x^{2}+y^{2},  \nonumber \\
s&=&\frac{(\frac{9}{2}\gamma_{m}^{2}-\frac{9}{2}\gamma_{m}+1)(1+x^{2}-y^{2})-3\sqrt{6}{\lambda}xy^{2}-10x^{2}+y^{2}-1}{3[(\frac{3}{2}\gamma_{m}-1)(1+x^{2}-y^{2})-2x^{2}-y^{2}-\frac{1}{2}]},  \nonumber \\
q&=&(\frac{3}{2}\gamma_{m}-1)(1+x^{2}-y^{2})-2x^{2}-y^{2},\label{rsxyph}
\end{eqnarray}
There is only one meaningful late-time attractor solution for
(\ref{phphaseplane}), which is the scalar field dominated
solution:
\begin{equation}
x=-\lambda/\sqrt{6},\qquad\qquad
y=\sqrt{1+\lambda^{2}/6}.\label{phsolution}
\end{equation}
and the eigenvalue of this solution is
$(-3-\lambda^2/2,-3\gamma_{m}-\lambda^2)$.

\begin{figure}
\begin{center}
\includegraphics[width=2.4in]{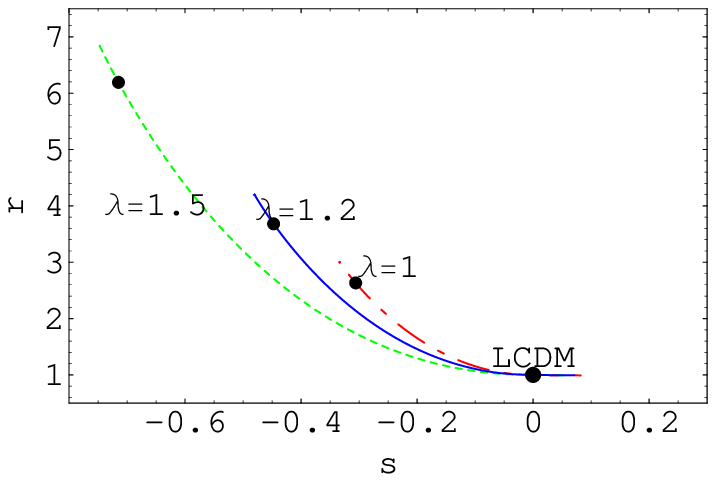}\includegraphics[width=2.4in]{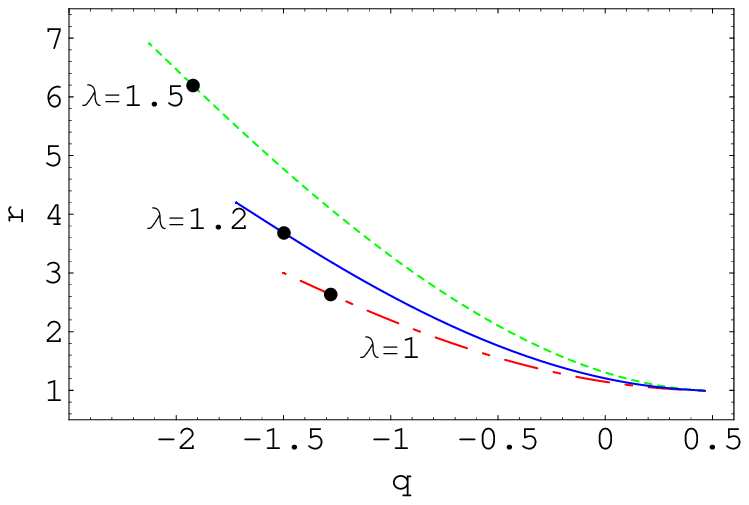}
\end{center}
\caption{The left figure is $s-r$ diagram of attractor solution in
phantom model. The right figure is $q-r$ diagram of attractor
solution in phantom model. The curves evolve in the variable
interval $N\in[-2,10]$. Selected curves for $\gamma_{m}=1$,
$\lambda=1$, $1.2$ and $1.5$ respectively. Dots locate the current
values of the statefinder parameters.}\label{sr2}
\end{figure}

\begin{figure}
\begin{center}
\includegraphics[width=2.4in]{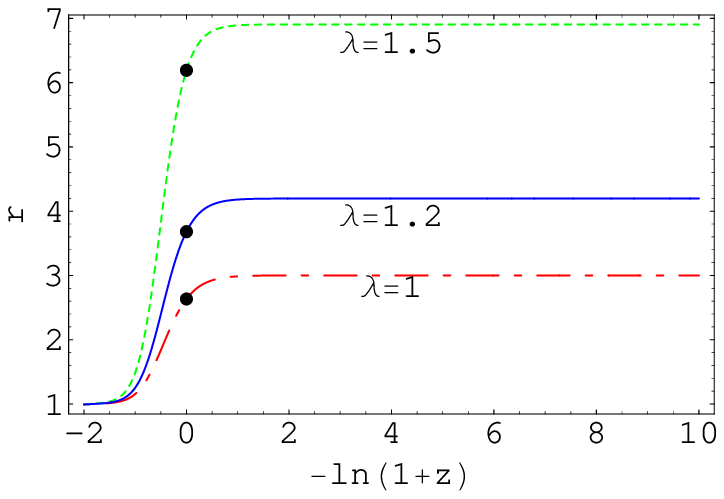}\includegraphics[width=2.4in]{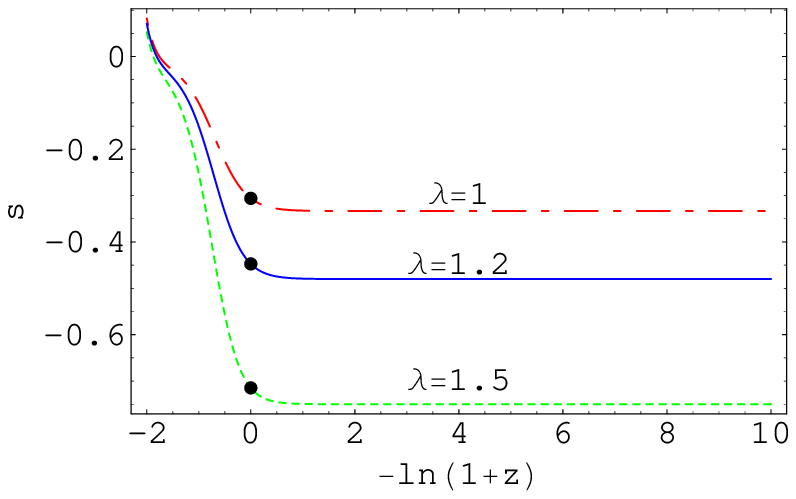}\includegraphics[width=2.4in]{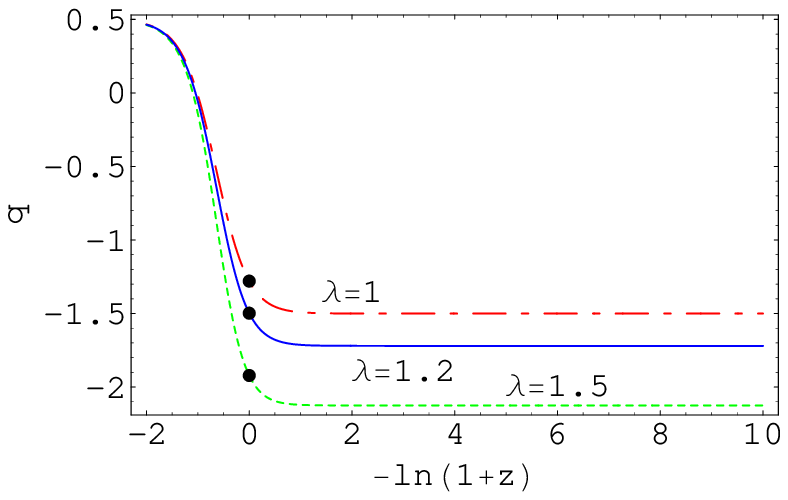}
\end{center}
\caption{The figures are the statefinder parameters and
deceleration parameter versus redshift diagram of the attractor
solution in phantom model. The left figure is $-\ln(1+z)$ --- $r$
diagram, the middle figure is $-\ln(1+z)$ --- $s$ diagram and the
right figure is $-\ln(1+z)$
--- $q$ diagram. The curves evolve in the variable interval $N\in[-2,10]$. Selected curves for $\gamma_{m}=1$,
$\lambda=1$, $1.2$ and $1.5$ respectively. Dots locate the current
values of the statefinder parameters.}\label{nr2}
\end{figure}

\begin{figure}
\begin{center}
\includegraphics[width=2.4in]{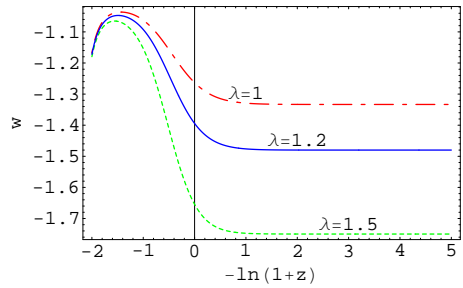}
\end{center}
\caption{The figures are the equation of state of the phantom
model $w$ versus redshift.  The curves evolve in the variable
interval $N\in[-2,5]$. Selected curves for $\gamma_{m}=1$,
$\lambda=1$, $1.2$ and $1.5$ respectively.}\label{phw}
\end{figure}

In Fig. \ref{sr2} we show the time evolution of the statefinder
pairs $\{r,s\}$ and $\{r,q\}$ for the attractor solution of the
phantom field. The plot is also for variable interval
$N\in[-2,10]$, and the corresponding parameters are the same in
the Fig. \ref{sr1}. We also can see that the curves far from the
LCDM in the future, but the statefinder pair $\{r,s\}$ lies in the
region $s<0$, $r>1$ which is different from that in Fig.
\ref{sr1}. We see that the statefinder trajectories $r-s$ is
almost linear in the past and future, which means that the
deceleration parameter changes from one constant to another nearly
with the increasing of the time, which can also be seen from the
Fig. \ref{phw}. In Fig. \ref{nr2}, the curves of $-ln(1+z)$
--- $r$ and $-ln(1+z)$ --- $s$ show clearly that the parameter $r$
will apart from $1$ and $s$ will far from $0$ with the increase of
the parameter $\lambda$. From $-ln(1+z)$ --- $q$ diagram we can
found that the deceleration parameter $q$ of phantom are all
satisfy $q<-1$ with different parameter $\lambda$ at present. In
Fig. \ref{phw} we see that the equation of state of phantom below
$-1$, which is different from
the Fig. \ref{quw}, and which could cause a big rip \cite{big rip,big rip1} in the universe in the future.\\
We have applied a statefinder analysis to the 5D attractor solutions
of quintessence and phantom field, and investigate the effect of the
parameter $\lambda$ in the potential $V(\phi)$. Though these two
attractor solutions are both scalar field dominated solution, the
difference of these two solutions can be found in the statefinder
parameters plane Fig.\ref{sr1}-Fig.\ref{phw}: (i) The region of the
statefinder pair $\{r,s\}$ is different between Fig. \ref{sr1} and
Fig. \ref{sr2}. For quintessence scenario Fig. \ref{sr1}, $s>0$ and
$r<1$ while for phantom scenario Fig. \ref{sr2}, $s<0$ and $r>1$.
(ii) The influence of the parameter $\lambda$ in potential $V(\phi)$
is different in the evolution of the universe, which can be seen in
the statefinder trajectories. In quintessence scenario Fig.
\ref{nr1}, the increasing of the $\lambda$ will decrease the
statefinder parameter $r$ and the deceleration parameter $q$, while
it will increase the statefinder parameter $s$. In phantom scenario
Fig. \ref{sr2}, the statefinder parameter $r$ will be increased,
while the $s$ and $q$ will be decreased with the increasing of the
$\lambda$ in the potential. So, through the statefinder parameter,
we can see the difference between quintessence field and phantom
field clearly. The attractor solution of quintessence field will
cause the universe to accelerate forever, but, the attractor
solution of phantom field will cause the universe to end with a big
rip. So, We think that we should pay attention to find a suitable
interaction between phantom and dark matter or other components in
5D model to avoid the big rip in our future work.

\section{Discussion}

In summary, we have studied the statefinder parameter $r$ and $s$
in the five-dimensional big bounce model in a plane-autonomous
system, and apply it to contrast the attractor solution of the
quintessence field and the phantom field respectively. It is found
that the evolving trajectories of these two attractor solutions in
the $\{r,s\}$ and $\{r,q\}$ plane is quite different, and which is
also different from the statefinder diagnostic of other dark
models. Through the figures of the statefinder parameters and
deceleration parameter we can see that the universe will
accelerate forever in the quintessence scenario, while the
universe will end with a big rip in the phantom scenario. We hope
that the future high precision observation will be able to
determine these statefinder parameters and consequently shed light
on the nature of dark energy.

\section{Acknowledgments}

This work was supported by NSF (10573003),  NBRP (2003CB716300) of
P. R. China and DUT 893321.

\end{document}